\documentclass[journal,comsoc]{IEEEtran}

\usepackage{setspace,amsmath,latexsym,cite,amssymb,epsfig,amsfonts}
\usepackage{url,cite}
\usepackage{graphicx}
\usepackage{psfrag}
\usepackage{footmisc}
\usepackage{multirow}
\usepackage{color}

\graphicspath{{.}{../eps/}{./images/}{./images/result/}}
\usepackage{amssymb}
\usepackage{amsthm}

\usepackage{cite}
\usepackage{mathtools, cuted}
\usepackage[font=small, labelsep=period]{caption}
\usepackage{makecell}
\usepackage{stfloats}
\usepackage{subcaption}
\usepackage{changepage}

\author{Peng~Xu,~\IEEEmembership{Member,~IEEE},
Gaojie~Chen,~\IEEEmembership{Senior Member,~IEEE,} Jianping Quan, Chong~Huang,~\IEEEmembership{Member,~IEEE,} Ioannis Krikidis, \IEEEmembership{Fellow, IEEE,}
Kai-Kit~Wong,~\IEEEmembership{Fellow,~IEEE}, and Chan-Byoung Chae,~\IEEEmembership{Fellow,~IEEE}
\thanks{
\noindent P. Xu and J. Quan are with the School of Communication and Information Engineering, Chongqing University of Posts and Telecommunications, Chongqing, 400065, China, and also with the Chongqing Key Laboratory of Mobile Communications Technology, Chongqing, 400065, China. (e-mail:xupeng@cqupt.edu.cn, quanjianp@163.com).

G. Chen is with School of Flexible Electronics (SoFE) \& State Key Laboratory of Optoelectronic Materials and Technologies, Sun Yat-sen University, Guangdong, China. (Email: gaojie.chen@ieee.org).

C. Huang is with the Institute for Communication Systems (ICS), Home for 5GIC \& 6GIC, University of Surrey, Guildford, Surrey GU2 7XH, U.K, and also with School of Flexible Electronics (SoFE) \& State Key Laboratory of Optoelectronic Materials and Technologies, Sun Yat-sen University, Guangdong, China. (E-mail: chong.huang@surrey.ac.uk).

I. Krikidis is with the Dept. of Electrical and Computer Engineering, University of Cyprus, Kallipoleos 75, Nicosia 1678, Cyprus. (Email: krikidis@ucy.ac.cy).

K. Wong is with the Department of Electronic and Electrical
Engineering, University College London, London WC1E 7JE, U.K. (Email:  kai-kit.wong@ucl.ac.uk). He is also affiliated with Yonsei Frontier Lab., Yonsei University, Seoul 03722 South Korea.

C.-B. Chae  is with the School of Integrated Technology, Yonsei University, Seoul 03722, South Korea. (Email: cbchae@yonsei.ac.kr).
}
}

\begin{document}

\title{\huge Deep Learning Driven Buffer-Aided Cooperative Networks for B5G/6G: Challenges, Solutions, and Future Opportunities}

\maketitle
{
\begin{abstract}

Buffer-aided cooperative networks (BACNs) have garnered significant attention due to their potential applications in beyond fifth generation (B5G) or sixth generation (6G) critical scenarios. This article explores various typical application scenarios of buffer-aided relaying in B5G/6G networks to emphasize the importance of incorporating BACN. Additionally, we delve into the crucial technical challenges in BACN, including stringent delay constraints, high reliability, imperfect channel state information (CSI), transmission security, and integrated network architecture. To address the challenges, we propose leveraging deep learning-based methods for the design and operation of B5G/6G networks with BACN, deviating from conventional buffer-aided relay selection approaches. In particular, we present two case studies to demonstrate the efficacy of centralized deep reinforcement learning (DRL) and decentralized DRL in buffer-aided non-terrestrial networks.
Finally, we outline future research directions in B5G/6G that pertain to the utilization of BACN.

\end{abstract}
}
\IEEEpeerreviewmaketitle
\section{Introduction} \label{sec:1}
\subsection{B5G/6G Vision and Background of BACN}

The beyond fifth generation (B5G) or sixth generation (6G) networks represent more than mere improvements or extensions of the 5G network; they signify remarkable paradigm shifts. The surge in mobile traffic is primarily driven by the rapid proliferation of new applications on mobile devices, including the Internet of things (IoT), the vehicle to everything (V2X), e-healthcare, machine-to-machine communications, and virtual/augmented reality. These applications demand increased network throughput, stringent network latency, and enhanced network reliability. As a result, besides interconnecting communication nodes, 6G will facilitate ubiquitous connectivity, ensure high-quality of service (QoS), and incorporate intelligent capabilities~\cite{6G_1}.

To meet the QoS requirements and serve as many devices as possible, a cooperative network has been introduced. The critical feature of cooperative transmission is to expand the coverage of the wireless network, enhance transmission reliability and increase user access. Besides, the relay and access points are commonly applied to enlarge the coverage of the terrestrial and non-terrestrial networks (NTNs). To further enhance the reliability, spectrum utilization and coverage of the system, the buffer-aided relaying was investigated as a promising technique in \cite{Zlatanov2014survey} and therein. The utilization of buffers in buffer-aided cooperative networks (BACNs) enables relays to temporarily store data packets and transmit them in two non-consecutive time slots, thereby providing greater flexibility in system scheduling and allowing for the optimization of network resources.

\begin{figure*}
    \centering
    \includegraphics[width=0.65\textwidth]{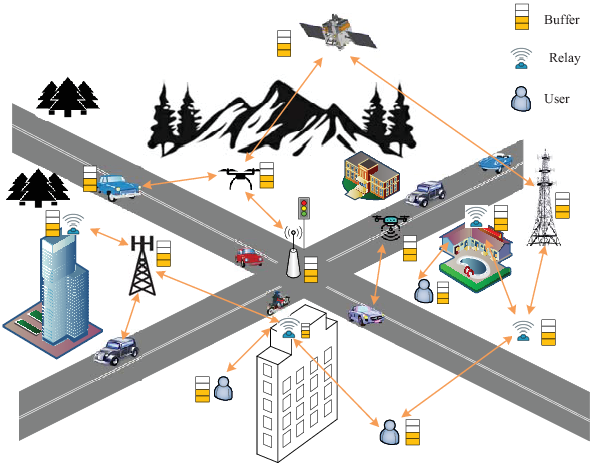}
    \caption{Communication scenarios of B5G/6G networks with buffer-aided relaying.}
    \label{fig1}
\end{figure*}

\subsection{Typical Applications}

{IoT systems connect not just human users but also a multitude of sensors within family homes or medical facilities. Buffer-aided relaying can store shared information at the dedicator in advance, which can help transmit data between two remote IoT nodes.
The V2X is a type of IoT that connects vehicles with surrounding entities such as roadside infrastructures, devices, pedestrians, grids, and nearby cars, which has become the backbone of the intelligent transportation system and tries to make the driving process safer and more intelligent\cite{Alam2020V2X}. The mobile vehicles can transmit urgent information to the central node, which can be stored in buffers and broadcast to new vehicles entering the intersection, as shown in Fig. \ref{fig1}.
However, V2X communication has high temporal-spatial dynamics, requiring buffers to update and replace information frequently due to the varying intrusions faced by vehicles at different locations and time slots.}



{The NTNs
provide wide coverage and mobile communication services through non-ground platforms such as satellites, high-altitude balloons, or unmanned aerial vehicles (UAVs). For satellite communications, the authors in \cite{yan2020satellite} investigated the relationship between QoS metrics, including buffer size, latency and transmit rate, and network configuration, by constructing a new tandem queuing model to imitate the data arrival, forwarding, and downloading processes in satellite data relay networks.
The authors in \cite{UAV-buffer2021} investigated a mobile relaying system assisted by a UAV with a finite buffer size, where the delay constraints are considered. Buffer-aided UAVs can dynamically adjust the NTN topology according to service requirements and carry data around obstacles if necessary. On the other hand, buffer-aided UAVs can ensure data transmission to the best of their ability through encountering severe weather because UAVs can store data packets in buffers, when the channel condition is weak, and transmit data packets to users once the channel gets better. Therefore, ad-hoc networks, including satellites and UAVs, can be applied to rescue operations in earthquake, flood or remote areas.}

\subsection{Relay Selection Schemes in BACNs}
{The traditional buffer-aided relay selection schemes are usually heuristically designed, considering both the wireless channel and buffer state.
For instance, the classical max-link method selects the link among all available links with the best channel
condition in each time slot, which achieves the full diversity order \cite{Krikidis2012constraints}.
However, since the max-link scheme only concerns the empty or full buffer state,
it has a large average packet delay that increases with the number of relays and
 the sizes of buffers. To address this, a buffer state-based relay selection scheme
 was proposed in \cite{luo2015BSB}, which considers both the channel quality and the
 buffer states.
Furthermore, to achieve high
  reliability and low delay, a deviation value-based relay selection scheme was proposed
  in \cite{XU2023dv}, where the allocation of deviation values depends on the disparity
  between instantaneous and predefined buffer state.}  

A favourable relay selection scheme design for BACNs needs to consider various aspects, including channel condition, buffer states, delay performance, etc. However, these aspects are time-varying with environmental change, and the delay requirements often conflict with outage requirements in the relay selection. For example, selecting the link with the best channel condition may not satisfy the buffer state requirement or cause a long delay. Solving these time-varying problems is still an open issue, especially when the buffer size or the relay number is significant. Therefore, the performance of the heuristically designed traditional buffer-aided relay selection schemes is far from the optimal bound.

This motivates researchers to investigate deep learning methods to solve relay selection in buffer-aided cooperative networks, which have advantages in solving high-dimensional and time-varying problems. Deep learning offers the advantage of being a data-driven tool, eliminating the need for pre-building mathematically model and allowing optimization based on a training data set. For example, in \cite{Huang2021IoT}, two deep reinforcement learning (DRL) based relay selection methods were proposed to solve a throughput maximization problem under strict delay constraints, which are different from most existing buffer-aided schemes usually considering average packet delay and achieve high throughput subject to delay constraints. However, incorporating deep learning algorithms into a BACN with multiple targets and constraints, such as secure transmissions, latency-aware transmissions, heterogeneous networks, and imperfect channel state information (CSI), is not a straightforward task. Furthermore, considering that the majority of current state-of-the-art deep learning techniques primarily focus on problems related to classification and natural language processing applications, the integration of complex targets and constraints into deep learning methods in general poses significant challenges.


\section{Technical Challenges Behind BACNs}
\label{sec:2}
Despite the apparent applications of BACNs in recent systems, there are several critical challenges in applying buffering to future wireless networks. The following section discusses some of these challenges.

\subsection{Low Latency}
To support the explosively growing demands of delay-sensitive wireless multimedia applications over the upcoming 6G wireless networks, the packet delay must be considered in BACN \cite{He2022security&delay}. Given a specific delay constraint, the design issues of QoS provisioning for multimedia wireless services have received considerable research attention. Because of the highly time-varying nature of wireless fading channels, deterministic delay-aware QoS constraints are no longer feasible to characterize queuing behaviours of multimedia wireless services. Towards this end, average and strict packet delay have been proposed to support delay-sensitive wireless communications over 6G multimedia mobile wireless networks. The existing works on buffer-aided cooperative communication mainly consider the average packet delay. However, the strict delay constraint in specific 6G scenarios is necessary, because the average packet delay can not ensure the delay fairness of different data packets. Still, the strict delay constraints can ensure the delay fairness of other data packets. Therefore, the pursuit of lower delay is a general trend in 6G, regardless of whether it is average or strict delay.

\subsection{High Reliability}
The trade-off between delay and reliability has also been investigated. With the increasing complexity of wireless systems and the number of nodes, balancing delay and reliability is a critical challenge. The expected massive number of devices of IoT in the network motivates 6G architecture to design a reliable transmission based on buffer-aided cooperative communication.  The future network must satisfy a high packet/data rate and reliable data communication between real-time applications (such as V2X, digital intelligent medical, etc.) and distributed edge devices. Specifically, the need for data reliability will be up to $99.9999\%$. 
The massive ultra-reliable low latency communication (mURLLC) services are required in 6G, and BACN faces reliability challenges when satisfying delay requirements.

\subsection{Transmission with Imperfect CSI}
Most existing works on BACN are developed based on instantaneous perfect CSI. However, obtaining instantaneous perfect CSI is difficult or impossible in large-scale networks due to the feedback delay and channel estimation errors \cite{Nomikos2018imperfect_CSI}. When accounting for the imperfect CSI, most existing traditional relay selection schemes have a high probability to select unqualified links for transmission, and thus the system performance will deteriorate, such as increased outage probability, reduced throughputs and longer delays. Consequently, designing new relay selection schemes with imperfect CSI is imperative, which is also challenging since the decider needs to evaluate the impact of the imperfect CSI on the system performance.

\subsection{Physical Layer Secure Transmission}
Transmitting a large amount of data, massive communication nodes, temporarily storing data in buffers, and the presence of potential eavesdroppers raise concerns about data security. In particular, due to the propagation characteristics of the wireless communication environment, information is most likely to be leaked during transmission.  Hence, appropriate buffer-aided relaying techniques and protocols shall be designed to protect user privacy and data security. However, owing to the existence of eavesdroppers,
the  buffer-aided relaying scheme  design is very challenging since some additional complicated factors
may need to be jointly considered, such as the   eavesdropping CSI, the secrecy throughput,
 and artificial noise, etc. In this case, a favourable  buffer-aided relaying scheme is hard to  design and
the analysis becomes very complex.  Moreover, with the  growth of the number of eavesdropping nodes in the
communication system, the heuristically designed buffer-aided relaying schemes are not easy to ensure secure transmission.

\subsection{Heterogeneous Network}
A heterogeneous network is a type of network that consists of different types of nodes and devices such as vehicles and some IoT nodes, which can support various services and applications. The integration of different networks provides seamless connectivity and increased coverage and capacity, enabling users to access services regardless of their location or device. Therefore, the heterogeneous network management to connect various nodes should be investigated. One of the main difficulties is managing the integration of different types of nodes and devices to provide seamless connectivity, increased coverage, and capacity. Additionally, the massive number of devices in a heterogeneous network leads to interference, which affects the network's performance. Therefore, buffer-aided relaying is an area that requires more investigation to improve the efficiency of data transmission in heterogeneous networks.
\vspace{1em}

\noindent
\emph{Note:} BACN also presents additional engineering challenges related to channel modeling, channel estimation, and the generation of realistic channel data. In demanding scenarios such as V2X, NTN, and drone communications, the channels exhibit high levels of dynamism, which poses difficulties in obtaining accurate channel information. Consequently, there is a crucial need for robust techniques that can efficiently generate and predict channel data with precision. One potential approach involves the fusion of ray-tracing and artificial intelligence algorithms, enabling the development of a super-resolution modeling method that satisfies the requirements for real-time simulation \cite{D.HE2023JSAC}.


\section{Potential Deep Learning Techniques} 
\label{sec:3}
\begin{figure*}[t!]
\begin{adjustwidth}{2em}{0em}
\begin{subfigure}[b]{.35\textwidth}
  \centering
  \includegraphics[width=1\linewidth]{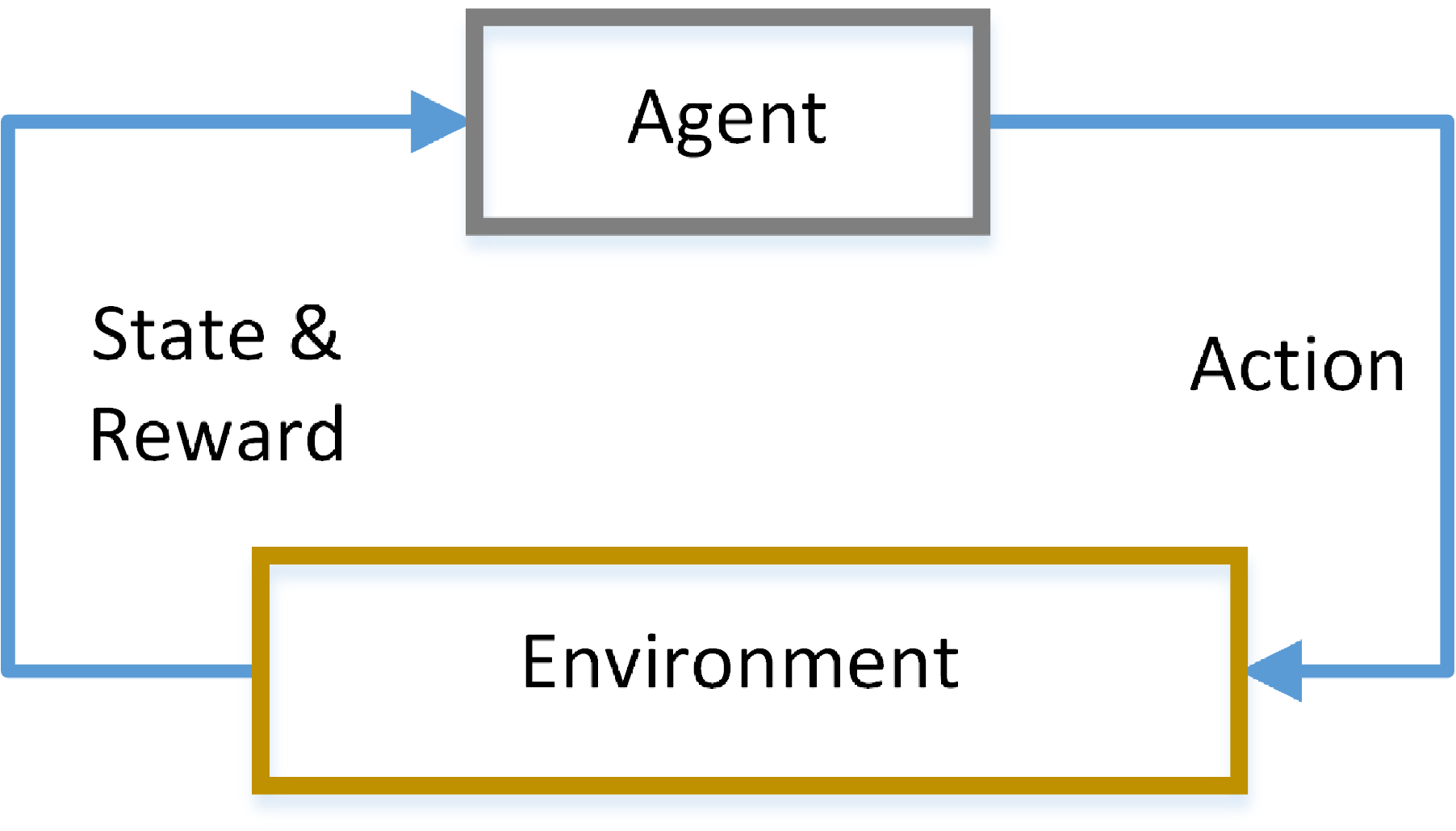}
  \caption{The framework of reinforcement learning. }
  \label{fig:dl3}
\end{subfigure}\hspace{30mm}
\begin{subfigure}[b]{.35\textwidth}
  \centering
  \includegraphics[width=1\linewidth]{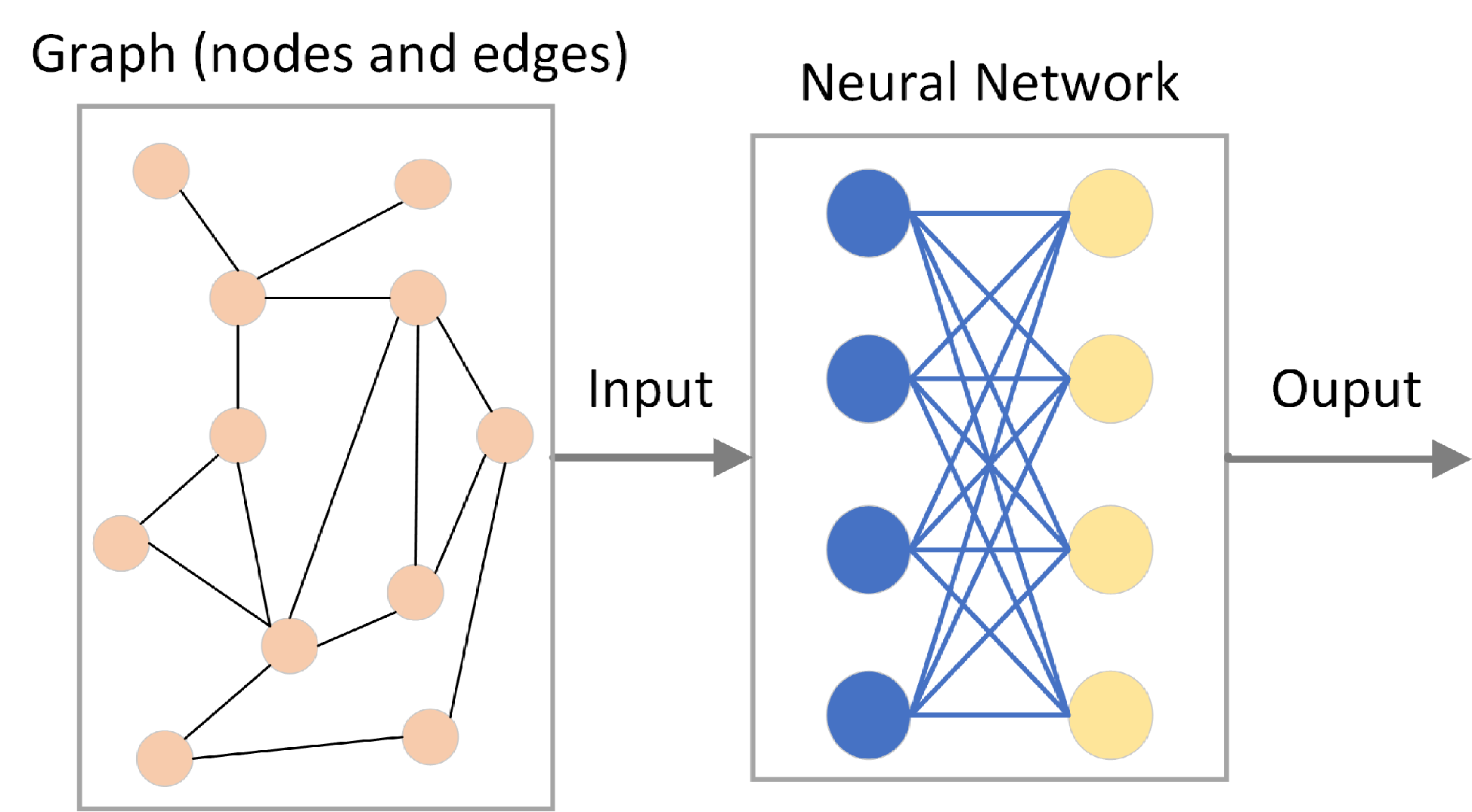}
  \caption{The structure of a graph neural network. }
  \label{fig:dl4}
\end{subfigure}
\end{adjustwidth}\vspace{3mm}
\begin{adjustwidth}{2em}{0em}
\begin{subfigure}[b]{.35\textwidth}
  \centering
  \includegraphics[width=1\linewidth]{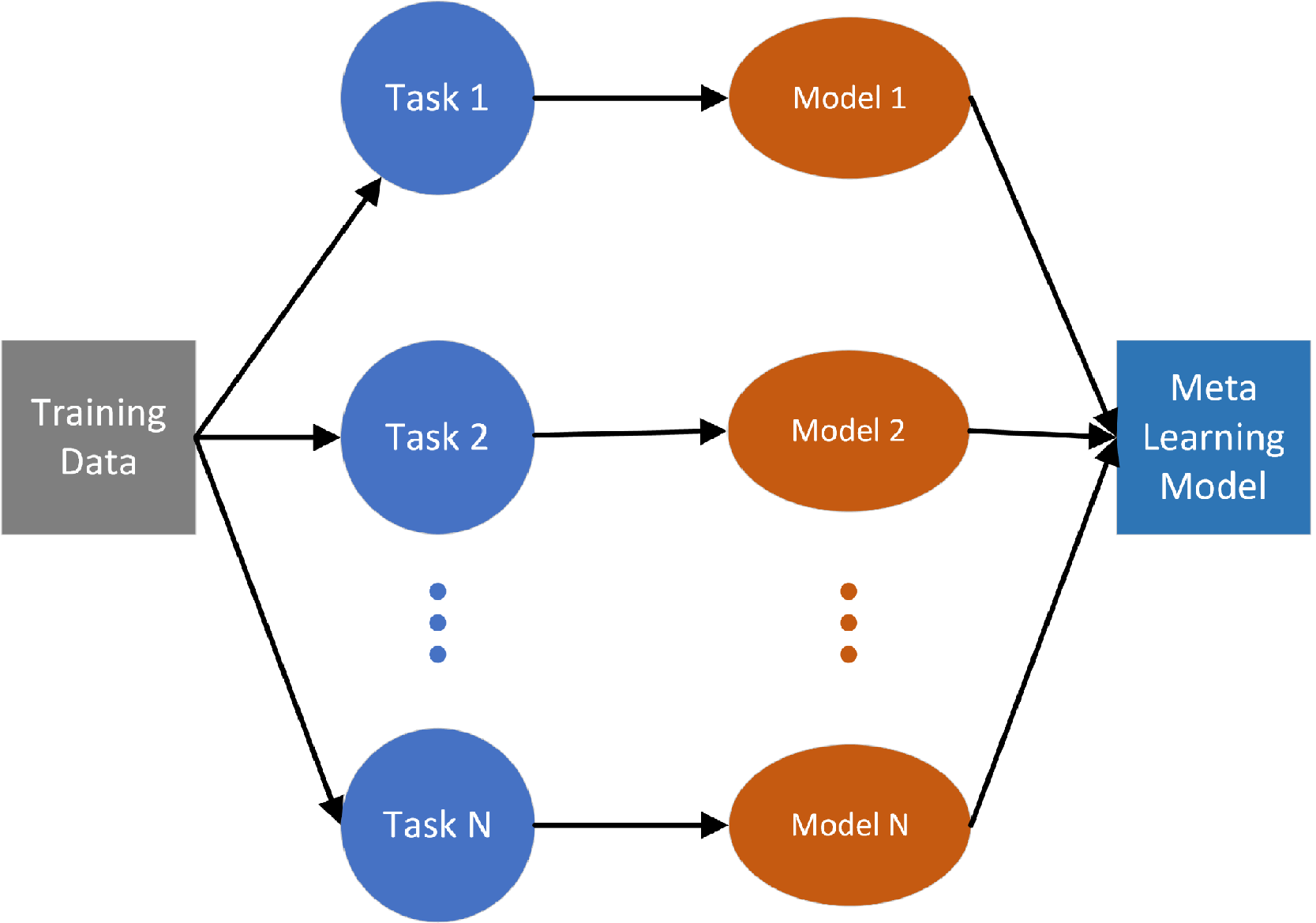}
  \caption{The structure of meta learning. }
  \label{fig:dl5}
\end{subfigure}\hspace{30mm}
\begin{subfigure}[b]{.35\textwidth}
  \centering
  \includegraphics[width=1\linewidth]{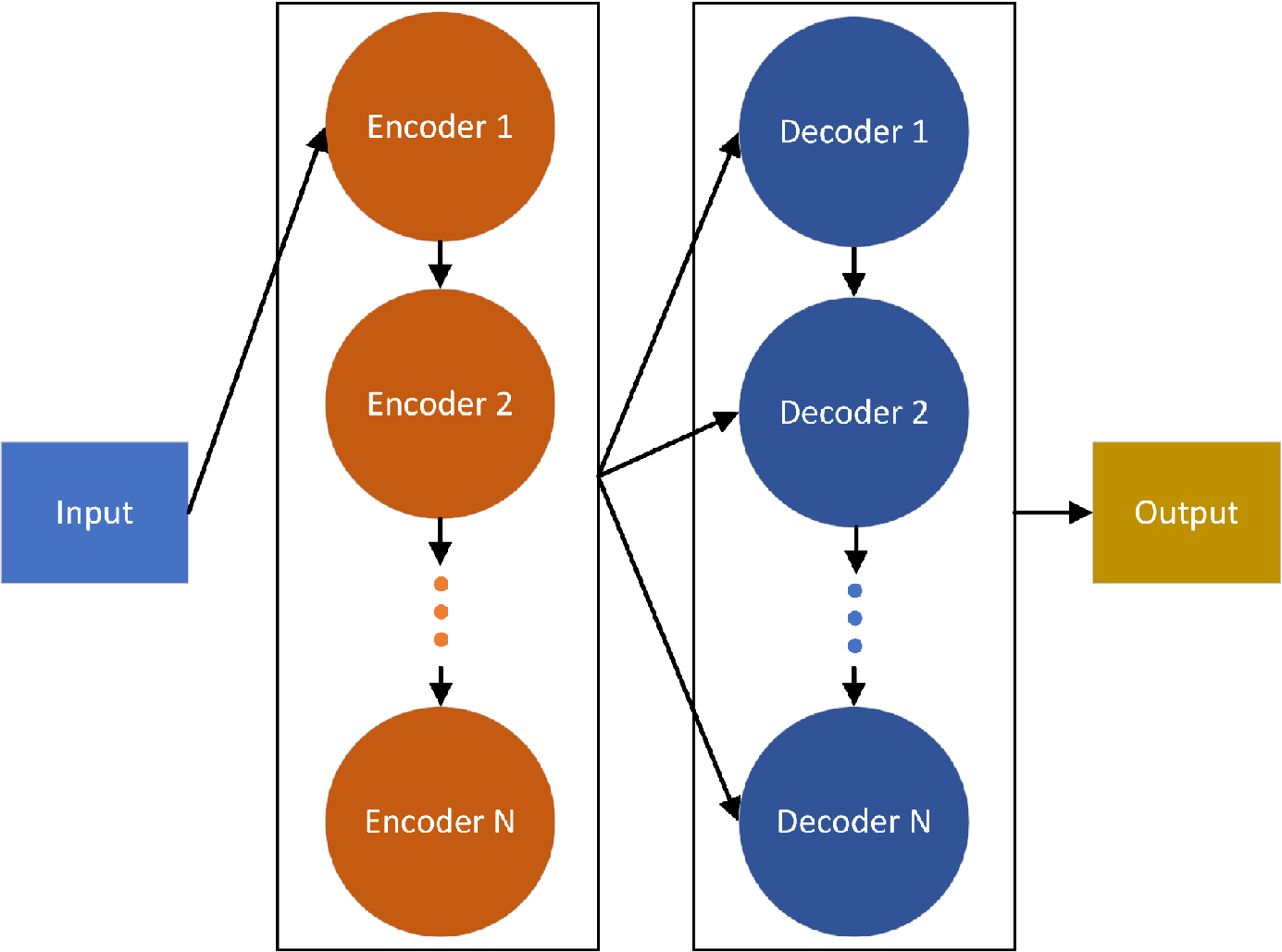}
  \caption{The structure of transformer learning. }
  \label{fig:dl6}
\end{subfigure}
\end{adjustwidth}
\caption{The structure of deep learning algorithms.}
\label{fig:dlfig}
\end{figure*}
In this section, current deep learning technologies that have the potential to be used in BACN are presented.

\subsection{Deep Reinforcement Learning}
A widely used solution in optimization problems of wireless communications is DRL. DRL algorithms are based on the idea of Markov Decision Processes (MDPs), which is a mathematical framework which uses states, actions, transitions and rewards to describe an environment, as shown in Fig. \ref{fig:dl3}. In MDPs, an agent is used to interact with the environment to learn the optimal decisions, this process is called reinforcement learning, as the agent learns from rewards received from the environment. Compared to traditional optimization methods and deep learning algorithms, DRL is more flexible to solve complex problems in dynamic environments such as BACNs.

\subsection{Graph Neural Network}
On the other hand, graph neural networks (GNNs) are a type of deep learning, specifically designed to process graph-structured data \cite{zhou2020graph}. In wireless communications, nodes in the graph represent devices in the network, and edges represent the relationships between devices, as shown in Fig. \ref{fig:dl4}. The matrix captures the relationships between different devices in BACNs and the neural network is used to process the related information to generate models to describe the relationships between devices in BACNs. Moreover, matrix and neural networks in GNN can process large amounts of data, which are very important in future large-scale BACNs.

\subsection{Meta Learning}
Recently, meta learning has been a widely used tool in the machine learning field \cite{vanschoren2019meta}. In traditional machine learning algorithms, a learning model is trained for a specific dataset to address a particular problem. However, the learned experience may not be suitable for new tasks anymore. Thus, meta learning is proposed to learn how to adjust the learning model from previous tasks to improve its adaptability to new tasks. In BACNs, meta learning can adjust the existing learning models for new tasks faster than training a new one. Therefore, meta learning can potentially improve the efficiency and flexibility of machine learning-based systems, enabling it to be important in future large-scale BACNs.

\subsection{Transformer}
Transformer is another important algorithm in current deep learning works. The key idea behind the transformer is self-attention, which allows it to simultaneously consider all positions of a sequence and learn contextual relationships between the tokens in the sequence. The Transformer model is utilized for learning long-term dependencies within the training dataset, this capability of sequential modeling enables it to infer the relationships between preceding and succeeding states more effectively in BACNs. Thus, it can optimize decision-making processes and enhances decision stability.

\subsection{Centralized and Distributed Learning}
In addition, centralized learning is a strategy for utilizing deep learning in wireless communications. Specifically, data from other devices in the network is collected in the control node and used to train a learning model for making decisions. The trained model in the control node is then used to make decisions and take actions for the whole network. This approach allows using large amounts of data to train a highly accurate model while avoiding transmitting large amounts of data between devices in the network during training.

On the other hand, in distributed learning, the computational resources of all devices in the network can be utilized, and the computational resources required at each device are reduced. Furthermore, distributed learning allows developing the optimization solution in large-scale networks, as each device can make decisions locally to reduce the amount of data transmitted in the network. Finally, distributed learning can improve privacy for the wireless network, as sensitive data does not need to be transmitted between devices.

\section{Proposed Frameworks for Deep Learning-Empowered BACNs} \label{sec:4}

\begin{figure*}
    \centering
    \includegraphics[width=0.75\textwidth]{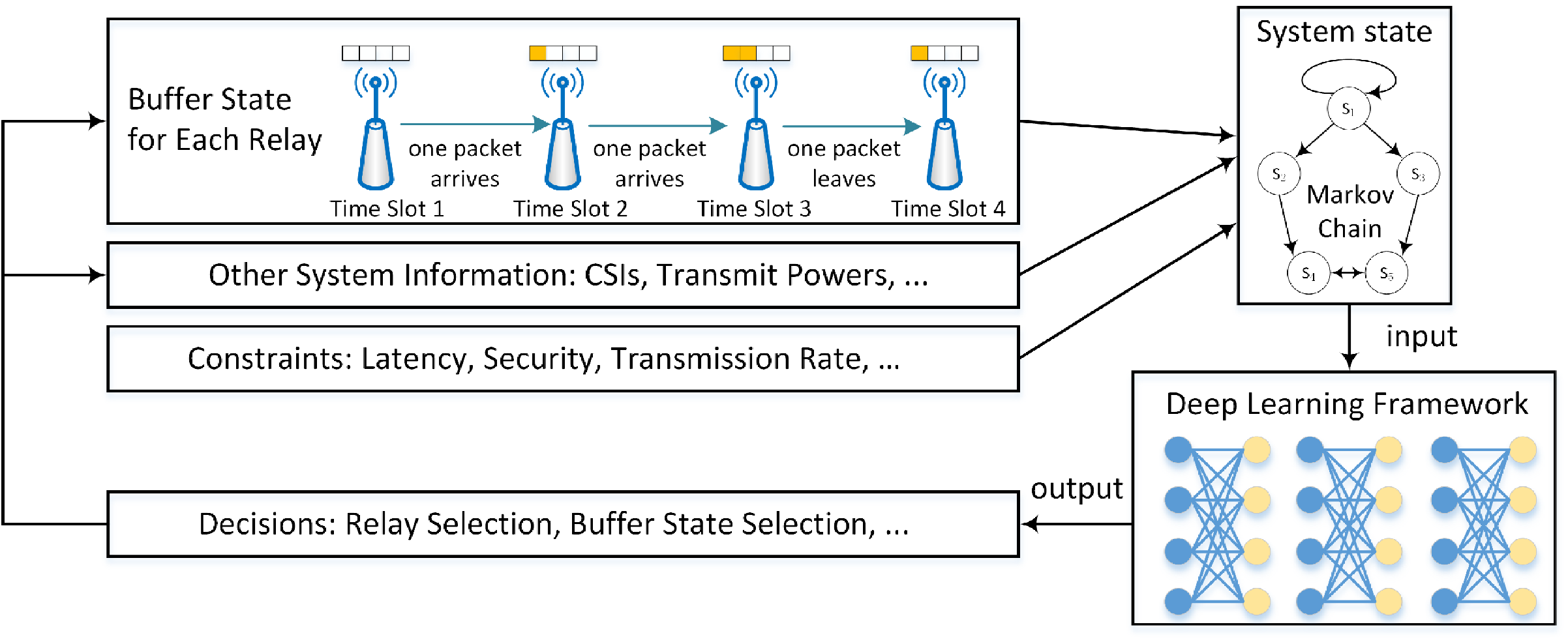}
    \caption{Deep learning framework for BACNs.}
    \label{fig_bacn_learning}
\end{figure*}
The main idea of deep learning is to use multiple layers of artificial neurons to learn the representation of data, one of the key advantages of using deep learning in wireless networks is that deep learning can learn the relationships between variables in the data, this allows deep learning to make accurate decisions about network optimization. In addition, deep learning algorithms can be trained by using historical data, which allows them to adapt to changes in the wireless network continuously. Moreover, the buffer-aided relay selection process can be modelled as an MDP \cite{HC2021JSAC}, and reinforcement learning algorithms can be applied to learn the solution from the MDP elements. Reinforcement learning focuses on mapping the situations to actions to maximize a reward. Thus, DRL algorithms can be utilized to optimize the tradeoff between the throughput and the delay in buffer-aided relay systems. Moreover, to improve the convergence performance of DRL in NTNs, meta-learning is a promising tool for adjusting the training model in DRL.

On the other hand, future wireless network could be a large-scale heterogenous system which leads to a more complex optimization problem. GNN is a promising tool for solving this problem by representing the graph structure as a matrix and processing this matrix by using a neural network. However, it is hard to obtain training data sets for GNN in NTNs. Thus, using GNN in combination with DRL can optimize the resource based on the decisions from DRL, where GNN help extract features of the network for training DRL.

As shown in Fig. \ref{fig_bacn_learning}, deep learning-based BACNs utilizes the power of deep learning algorithms to optimize the resource in BACNs. The system state information, which includes buffer state, CSI and optimization constraints, has the potential to transition to multiple states in the subsequent time period, thus forming a Markov chain. Besides, the system state is regarded as the input of the deep learning framework, which is designed as the optimizer in BACNs. The deep learning algorithms mentioned above is combined and used to form the deep learning framework to train a neural network model for making decisions in BACNs. The pre-trained model helps the deep learning framework to optimize the outputs as the decisions variables for BACNs, such as buffer-aided relay selection and routing strategy.

\begin{figure*}[t!]
\begin{adjustwidth}{0em}{0em}
\begin{subfigure}[b]{.45\textwidth}
  \centering
  \includegraphics[width=1\linewidth]{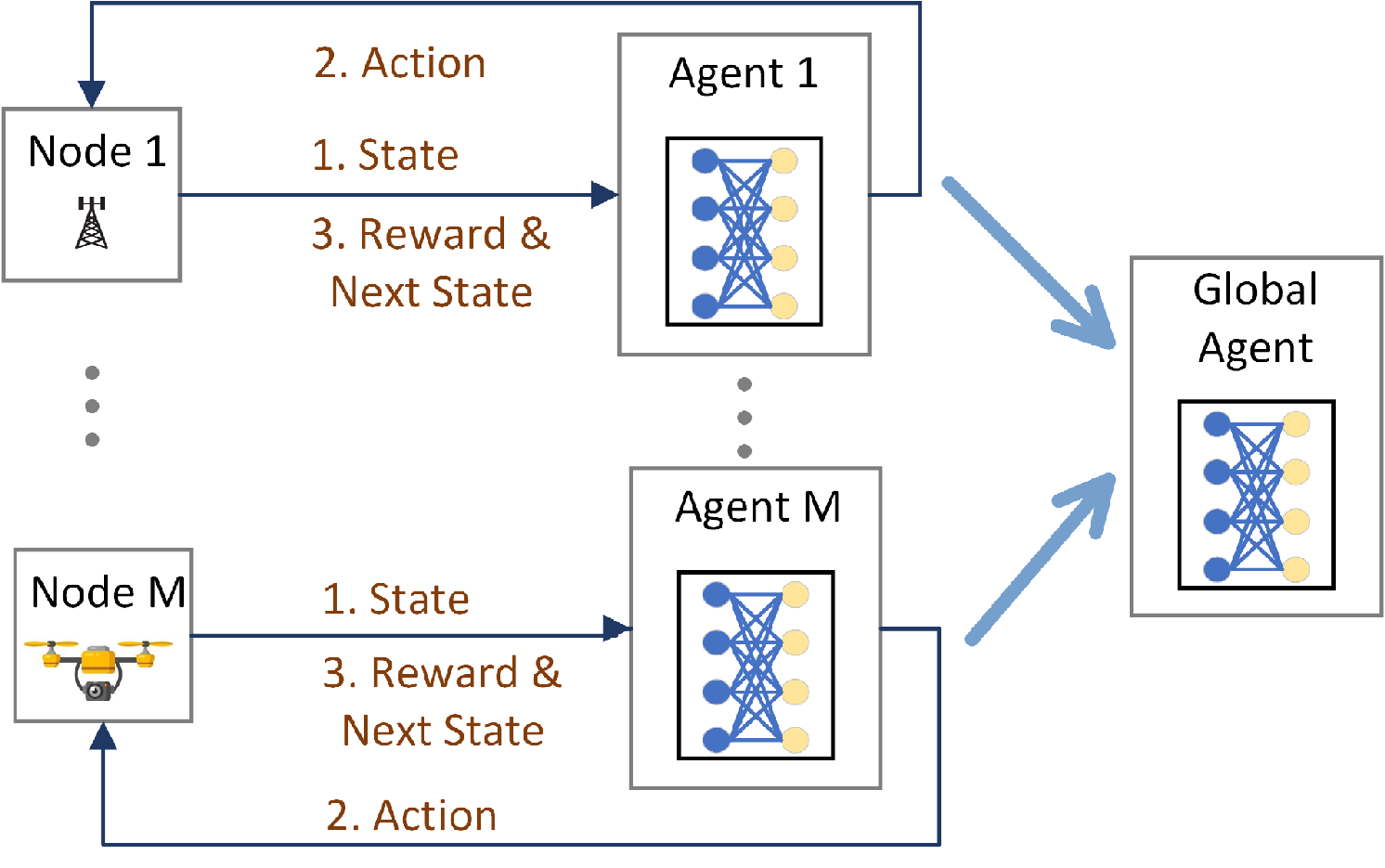}
  \caption{Decentralized DRL in NTN. }
  \label{fig:sim_decen}
\end{subfigure}\hspace{10mm}
\begin{subfigure}[b]{.45\textwidth}
  \centering
  \includegraphics[width=1\linewidth]{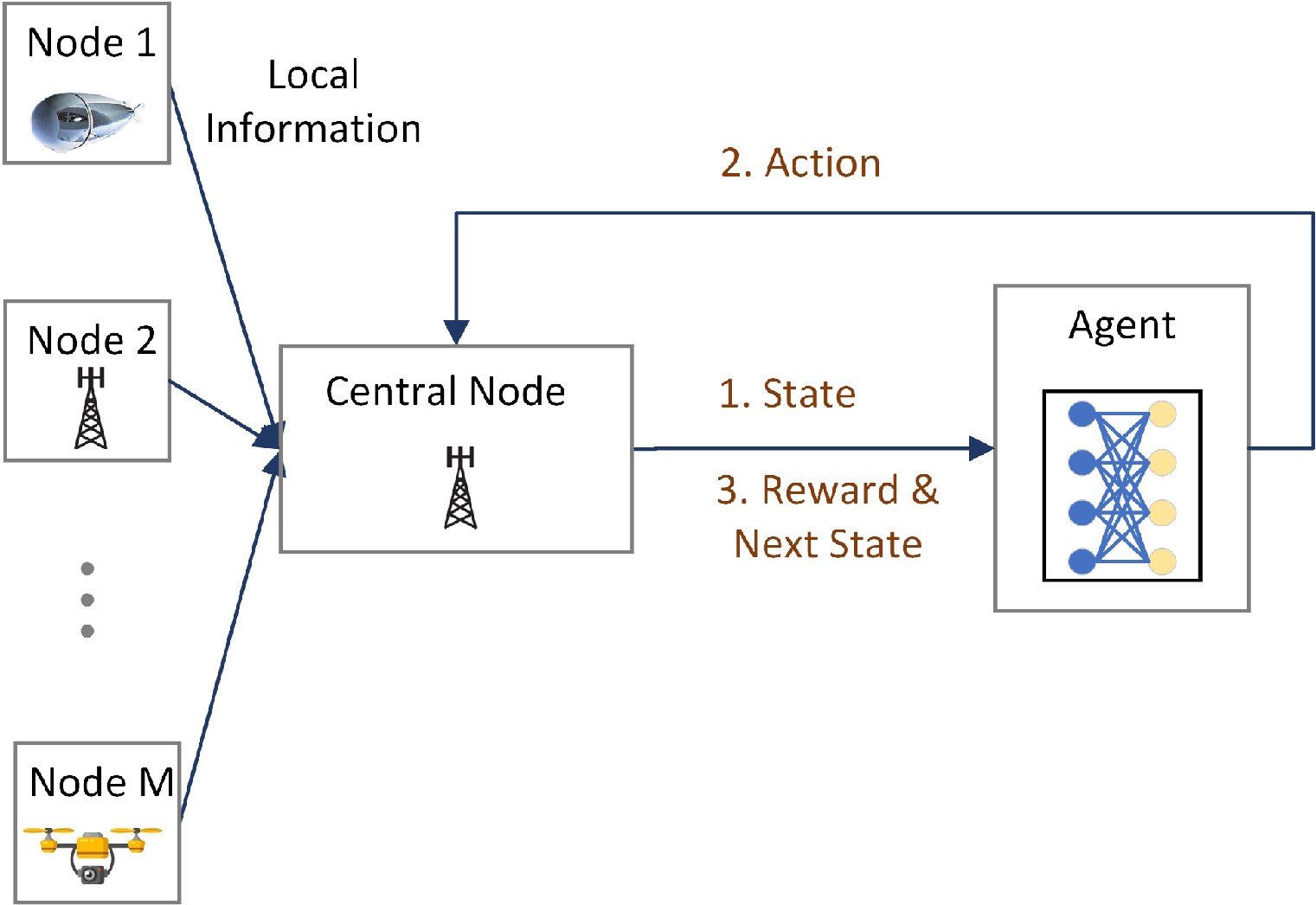}
  \caption{Centralized DRL in NTN.}
  \label{fig:sim_cen}
\end{subfigure}
\end{adjustwidth}
\caption{The structure of decentralized DRL and centralized DRL in non-terrestrial networks (NTNs) with buffer relaying.}
\label{fig:sim_cen_decen}
\end{figure*}

Moreover, most of the existing buffer-aided relay selection schemes assume that there is a central control node to receive all required information in the network and make global decisions. However, considering the communication cost, the relay nodes may only receive the required information from neighbour nodes, or the limitations of the hardware cannot allow the central node to receive all information and calculate a solution. Therefore, a DRL algorithm, multi-agent proximal policy optimization (MAPPO), is introduced to solve this problem. In that case, each agent works on a communication node, and all the agents could learn to obtain a solution through their interactions and the local environment. Then all agents cooperate in a distributed way to improve the sum throughput with delay constraints. Moreover, GNN is utilized to model network graphs to captures the relationships between different devices in the proposed network to help optimize DRL decisions. Considering the outdated CSI in the proposed NTN and the high convergence complexity in MAPPO, meta-learning is embedded in the proposed MAPPO algorithm to adjust the training model. Specifically, meta-learning can learn the relationship between the previously trained models and their predictions and then build a meta-model for adjusting the hyperparameters in the decision model.

To give the performance of the deep learning-based optimization in buffer-aided NTNs, we assume a low Earth orbit satellite as the source, multiple HAPs, UAVs and base stations as the relay nodes, and a ground user as the destination. Each relay node is equipped with a buffer and operates in half-duplex mode. Thus, a relay node cannot transmit and receive simultaneously in the multi-hop network. All channels are assumed to experience Rician fading. An untrusted user is assumed to be an eavesdropper who can eavesdrop on the signals in the buffer-aided NTN to present the challenge of physical layer secure transmission. Besides, the strict delay constraint is considered in the throughput performance to present the challenge of low latency. Moreover, the outdated CSI in the proposed network brings the challenge of imperfect CSI in wireless transmissions. In the face of these constraints and challenges, it is crucial to leverage the proposed deep learning framework to maximize system throughput and ensure reliable transmission.

\begin{figure}[t!]
  \centering
  \centerline{\includegraphics[scale=0.5]{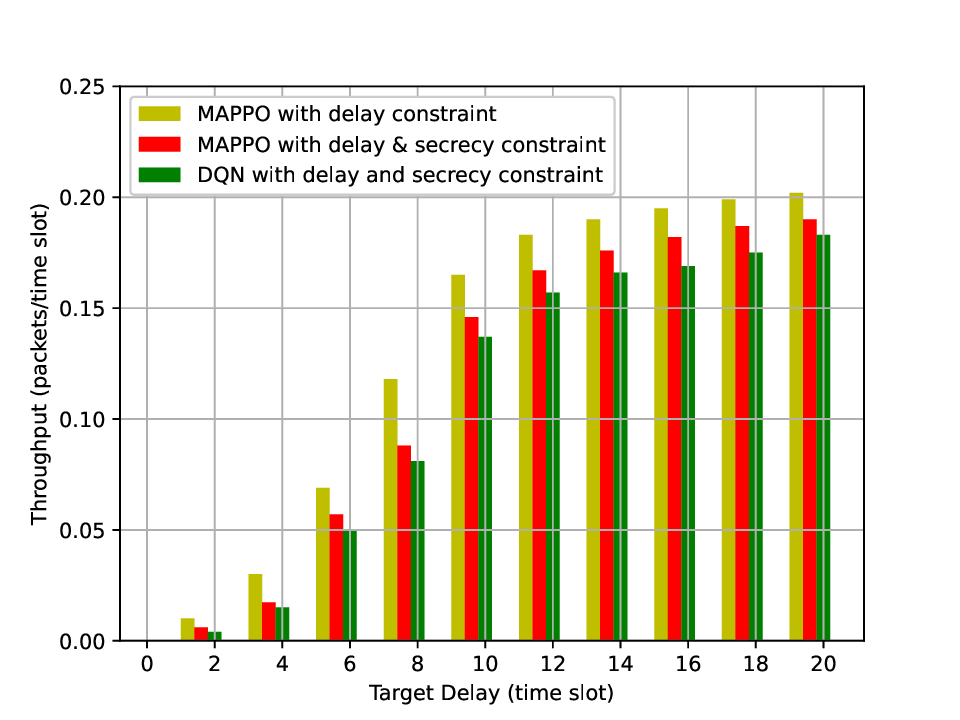}}
 \caption{ Throughput {\em vs.} target delay in decentralized NTN with buffer relaying.} \label{fig:result2}
\end{figure}

\begin{figure}[t!]
  \centering
  \centerline{\includegraphics[scale=0.5]{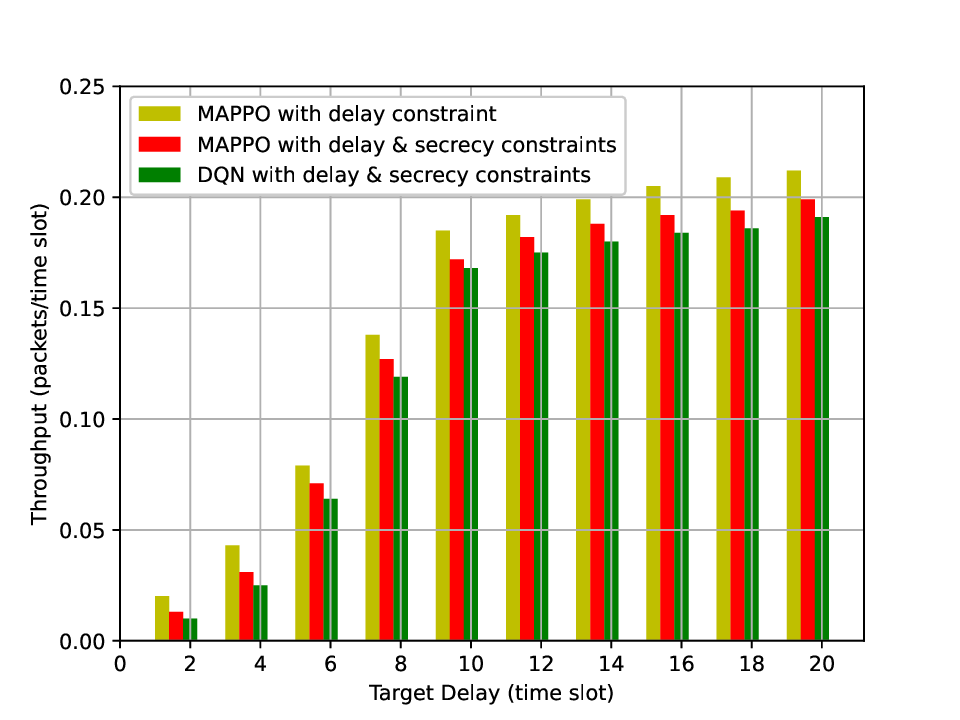}}
 \caption{ Throughput {\em vs.} target delay in centralized NTN with buffer relaying.} \label{fig:result1}
\end{figure}

Considering that central control is infeasible in a decentralized NTN, we employ MAPPO as shown in Fig. \ref{fig:sim_decen} to optimize the multi-hop routing strategy in the NTN with buffer relaying by interacting with the MDP environment for 5000 time slots. In Fig. \ref{fig:result2}, it is clear that although the presence of the eavesdropper decreases the performance of throughput, the proposed MAPPO-based algorithm still can optimize the solution to maximize the throughput with delay and secrecy constraints. Besides, considering each agent can only obtain knowledge from adjacent nodes, all agents learn from their local environment and cooperate to build a joint solution. Moreover, compared to a deep Q-Learning network (DQN), MAPPO could converge better by using the improved trust region update strategy.

Fig. \ref{fig:result1} shows the performance of DRL algorithms in a centralized buffer-aided NTN. In the centralized scenario, we assume a central controller which could be any node which is strategically positioned to receive global information optimally in the practical network. The structure of the proposed centralized DRL algorithm is shown in Fig. \ref{fig:sim_cen}. The results indicate that the target delay strongly impacts the throughput with delay constraints in the NTN. Buffer technology improves the transmission reliability for low signal-to-noise ratio (SNR) channels, but using a buffer usually leads to the cost of delay. The proposed MAPPO-based algorithm could interact with the environment to determine the impact of the delay constraint and learn to find the solutions for each system state. Thus, the MAPPO-based algorithm could optimize the solution with different target delays. Besides, considering the presence of an eavesdropper, the MAPPO-based algorithm needs to consider the secrecy rate while choosing relays for transmission. Thus, the performance slightly decreases because the agent may avoid selecting relay nodes around the eavesdropper. In addition, the centralized approach outperforms the distributed one under various delay constraints. The reason is that in a centralized wireless communication environment, a central controller has access to all information to achieve enhanced optimization results through learning from the comprehensive data. In contrast, in a distributed setting, it is challenging to transmit information from every node to a central node for optimal global control. Thus, each node has to optimize the strategy based on its judgement and information received from adjacent nodes, resulting in suboptimal performance compared to centralized systems.

\section{Future Research Directions}
\label{sec:5}

Although BACN in 6G communication is an exciting area of interest for the research community, the research on relay selection and buffer state management in BACN is still in its early stages. Therefore, we highlight the main research directions for adopting buffer-aided relaying into 6G networks.

\subsection{Future Deep Learning in BACN}
In the future, deep learning will play an important role in BACN. However, considering BACN will be more complex (such as massive relay nodes, the dynamic feature of relay nodes' appearing and disappearing, and energy consumption at relays) and connected to other networks such as satellite networks in the future 6G communications, existing deep learning methods may suffer from the high difficulty of generating training data set and large decision space of the system. Thus, some other new technologies in deep learning could be introduced in BACN to address these challenges for future 6G networks. For example, the double cascade correlation network can be utilized to generate training data set, while Wolpertinger architecture is designed to reduce the cost of evaluating actions in MDPs. Moreover, considering the BACN scenarios can be changed in real communication systems, to build an adaptive decision system for BACN, the metaheuristic algorithm which can further optimize the deep learning-based decision systems could be utilized to improve the convergence performance for different BACN scenarios.

\subsection{Compatible with the Reconfigurable Intelligent Surface}
The buffer-aided relaying could also connect with the reconfigurable intelligent surface (RIS) to improve the wireless environment. Beyond the inherent challenges behind BACNs, there are some new challenges. Firstly, optimizing the performance of complicated channels, such as those involving RIS to active BSs/buffer-aided relays in hybrid wireless networks, is a critical issue that needs to be addressed. Secondly, designing the passive beamforming elements for RIS contains discrete amplitude and phase-shift levels. The exhaustive search is impractical; some existing works relax constraints and obtain the approximate values. However, it may lead to performance loss due to quantization errors. To further improve the performance, the heuristic alternating optimization technique can be applied to iteratively optimize the discrete amplitude/phase-shift values. Then the optimized results can be utilized as training data set for deep learning-based methods. Moreover, considering the computational complexity of heuristic alternating optimization and DRL methods for practical RIS coefficients, the double cascade correlation network can be introduced to speed up the convergence and improve the performance.

\subsection{Security and Privacy Issues}
For buffer-aided relaying, existing works mainly focus on efficiently transmitting data. However, some users may refuse to assist other users in transmitting information due to privacy concerns about sharing sensitive information that could result in a security breach or technical difficulties. The privacy and security have gained more and more attention in B5G/6G, where users may need to keep their data locally and prevent eavesdropping. On the other hand, training data in the artificial intelligence model should also attach security issues. Therefore, the distributed training architecture must fully consider the data allocation, computing capability, and network. Thus, the new machine-learning-driven buffer-aided relay selection methods, e.g. federated learning, can assure privacy and security in demand. Moreover, deep learning algorithms can be used to encrypt data, making it more difficult for attackers to intercept and access sensitive information in BACN.

\subsection{Hybrid Buffer and Cache Selection Technology}
Wireless caching is an emerging technology that enables storing popular content in caches, allowing for direct transmission from the cache instead of the remote cloud. To further enhance transmission efficiency, buffer-aided selection schemes can be combined with caching technology. Deep learning algorithms can be employed to tackle the challenges associated with the design and optimization of such hybrid buffer and cache schemes, which possess the ability to handle complex and high-dimensional data and the potential for discovering intricate patterns and relationships within the data. By leveraging deep learning algorithms, hybrid buffer and cache selection technologies can be developed to efficiently manage content delivery in 6G networks.
\subsection{BACN in Future 6G SAGIN}
The space-air-ground integrated networks (SAGINs) contain a large number of dynamic nodes in their wide connection range, the buffer-aided relay selection in SAGIN is promising. Firstly, a future direction is to deal with the complex nature of dynamic devices, such as UAVs and satellites in SAGIN. Some existing works have investigated buffer-aided relaying in UAV networks. However, the rapid mobility issue of UAVs remains unresolved. On the other hand, communication constraints among different communication layers and high overhead exist caused by the highly dynamic environment of SAGIN are critical challenges. In addition, it is difficult to obtain channel information in a dynamic environment. Therefore, the deep-learning-driven methods such as inverse reinforcement learning for buffer-aided relaying represents may effectively manage resource and channel acquisition for flexible SAGIN.

\section{Conclusions}
\label{sec:6}

 This article has highlighted the potential for improvement in existing relay selection schemes for buffer-aided cooperative systems to meet emerging demands. The challenges associated with using BACNs have been identified, and it has been demonstrated that deep learning-based solutions hold promise in addressing these challenges. To achieve secure communication with ultra-reliable and low latency requirements, considering imperfect CSI and heterogeneous networks, a clear understanding of the intrinsic characteristics of BACN is crucial. In contrast to conventional buffer-aided relay selection methods, this article has proposed deep learning-driven approaches applicable to both centralized and decentralized scenarios. Through simulation results, the performance of DRL has been demonstrated in a centralized buffer-aided NTN, along with the effectiveness of the multi-agent DRL algorithm in a decentralized NTN. The proposed deep learning-driven methods have the potential for generalization across various BACNs and offer valuable insights for future research directions.

\bibliographystyle{IEEEtran}
\bibliography{reference}
\begin{IEEEbiographynophoto}{Peng Xu}(Member, IEEE) received the B.Eng. and the Ph.D. degrees in electronic and information engineering from the University of Science and Technology of China, Anhui, China, in 2009 and 2014, respectively. From June 2014 to July 2016, he was  working as a postdoctoral researchers with the Department of Electronic Engineering and Information Science, University of Science and Technology of China, Hefei, China. He is currently an Associated Professor with the School of Communication and Information Engineering, Chongqing University of Posts and Telecommunications (CQUPT), Chongqing, China. His current research interests include cooperative communications, information-theoretic secrecy, NOMA techniques and reconfigurable intelligent surface. Dr. Peng Xu received {\scshape IEEE Wireless Communications Letters} Exemplary Reviewer (2015 and 2021) and Excellent Paper of Chongqing Association for Science and Technology (2018).
\end{IEEEbiographynophoto}

\begin{IEEEbiographynophoto}{Gaojie Chen} (Senior Member, IEEE) received PhD degree from Loughborough University, U.K. in 2012. After graduation, he took up academic and research positions at DT Mobile, Loughborough University, the University of Surrey, the University of Oxford and the University of Leicester. His research interests include wireless communications, satellite communications and secrecy communication. He is currently a Professor at the School of Flexible Electronics (SoFE) \& State Key Laboratory of Optoelectronic Materials and Technologies, at Sun Yat-sen University. He serves as an Associate Editor for the {\scshape IEEE Transactions on Wireless Communications}, the {\scshape IEEE Communications Letters} and {\scshape IEEE Wireless Communications Letters}.
\end{IEEEbiographynophoto}

\begin{IEEEbiographynophoto}{Jianping Quan}(Student Member, IEEE) received the B.Sc. degree in communication engineering at the School of Communication and Information Engineering, Chongqing University of Posts and Telecommunications (CQUPT), Chongqing, China in 2020, where she is currently pursuing the M.Sc. degree in information and communication engineering. Her research interests include cooperative communication, NOMA techniques, the simultaneous wireless information and power transfer (SWIPT) system and machine learning.
\end{IEEEbiographynophoto}

\begin{IEEEbiographynophoto}{Chong Huang} (Member, IEEE) received the BEng degree in communication engineering from Nanjing University of Posts and Telecommunications in 2011, the MSc degree in electrical and electronic engineering from Loughborough University in 2015 and the Ph.D. degree in wireless communications with the University of Surrey in 2023. He is currently an Research Fellow at the 6GIC, University of Surrey. His research interests include reinforcement learning, cooperative networks, physical layer security, cognitive radio, non-orthogonal multiple access, reconfigurable intelligent surfaces, edge computing, satellite communications and Internet of Things.
\end{IEEEbiographynophoto}

\begin{IEEEbiographynophoto}
{Ioannis Krikidis}
(Fellow, IEEE) received the Diploma degree in computer engineering from the Computer Engineering and Informatics Department
(CEID), University of Patras, Greece, in 2000, and the M.Sc. and Ph.D. degrees in electrical engineering
from the \'{E}cole Nationale Sup\'{e}rieurennnn des T\'{e}l\'{e}communications (ENST), Paris, France, in 2001 and 2005, respectively. He is currently an Associate Professor at the Department of Electrical and Computer Engineering, University of Cyprus, Nicosia, Cyprus. He serves as an Associate Editor for the IEEE TRANSACTIONS ON WIRELESS COMMUNICATIONS, and IEEE WIRELESS COMMUNICATIONS LETTERS. He has been recognized by the Web of Science as a Highly Cited Researcher, 2017-2021. He has received the prestigious ERC Consolidator Grant.
\end{IEEEbiographynophoto}

\begin{IEEEbiographynophoto}
{Kai-Kit Wong}(Fellow, IEEE)
received the B.Eng., M.Phil., and Ph.D. degrees in electrical and electronic engineering from The Hong Kong University of Science and Technology, Hong Kong,
in 1996, 1998, and 2001, respectively. He is Chair in Wireless Communications in the Department of Electronic and Electrical Engineering, University College London, United Kingdom. He is a Fellow of IET and is also on the Editorial Boards of several international journals. He has been the Editor-in-Chief of IEEE Wireless Communications Letters since 2020.
\end{IEEEbiographynophoto}

\begin{IEEEbiographynophoto}
{Chan-Byoung Chae}(Fellow, IEEE)
received the Ph.D. degree in electrical and computer engineering from The University of Texas at Austin (UT) in 2008. Prior to joining UT, he was a Research Engineer at the Telecommunications Research and Development Center, Samsung Electronics, Suwon, South Korea, from 2001 to 2005. He is currently an Underwood Distinguished Professor with the School of Integrated Technology, Yonsei University, South Korea. Before joining Yonsei University, he was with Bell Labs, Alcatel-Lucent, Murray Hill, NJ, USA, from 2009 to 2011, as a Member of Technical Staff, and Harvard University, Cambridge, MA, USA, from 2008 to 2009, as a Post-Doctoral Research Fellow. He was an EiC of IEEE Trans. Molecular, Biological, and Multi-scale Communications.


\end{IEEEbiographynophoto}

\end{document}